\documentclass[10pt,journal,compsoc]{IEEEtran}
\ifCLASSOPTIONcompsoc
  \usepackage[nocompress]{cite}
\else
  \usepackage{cite}
\fi
\ifCLASSINFOpdf
   \usepackage[pdftex]{graphicx}
   \graphicspath{{pictures/}}
\else
\fi
\usepackage{amsmath}
\usepackage{url}

\usepackage{xcolor}

\begin{document}
%
\title{The "Kinesthetic HMD":\\ 
Enhancing Self-Motion Sensations in VR\\ with Head-Based Force Feedback.}
%
%
%
%

\author{Antoine~Costes, 
        Anatole~L\'ecuyer
\IEEEcompsocitemizethanks{\IEEEcompsocthanksitem Antoine Costes was with Inria, Univ Rennes, CNRS, IRISA, France.\protect\\
E-mail: see http://www.antoinecostes.net
\IEEEcompsocthanksitem Anatole L\'ecuyer was with Inria, Univ Rennes, CNRS, IRISA, France.\protect\\
E-mail: anatole.l\'ecuyer@inria.fr}
\thanks{Manuscript received April 19, 2005; revised August 26, 2015.}}

%
%

\markboth{Journal of \LaTeX\ Class Files,~Vol.~14, No.~8, August~2015}%
{Costes and L\'ecuyer: The "Kinesthetic HMD": Inducing illusory whole-body motion in immersive virtual reality with head-based force feedback.}
%



\IEEEtitleabstractindextext{%
\begin{abstract} 

The sensation of self-motion is essential in many virtual reality applications, from entertainment to training, such as flying and driving simulators.
If the common approach used in amusement parks is to actuate the seats with cumbersome systems, multisensory integration can also be leveraged to get rich effects from lightweight solutions.
In this short paper, we introduce a novel approach called the "Kinesthetic HMD": actuating a head-mounted display with force feedback in order to provide sensations of self-motion. 
We discuss its design considerations and demonstrate an augmented flight simulator use case with a proof-of-concept prototype.
We conducted a user study assessing our approach's ability to enhance self-motion sensations. 
Taken together, our results show that our Kinesthetic HMD provides significantly stronger and more egocentric sensations than a visual-only self-motion experience.
Thus, by providing congruent vestibular and proprioceptive cues related to balance and self-motion, the Kinesthetic HMD represents a promising approach for a variety of virtual reality applications in which motion sensations are prominent. 
\end{abstract}

}

\maketitle

\IEEEdisplaynontitleabstractindextext

%
\IEEEpeerreviewmaketitle

\IEEEraisesectionheading{\section{Introduction}\label{sec:introduction}}

%
%
%
%
\IEEEPARstart{W}{ith} the blossoming of virtual reality (VR) consumer-grade devices in the last five years, numerous applications for VR bloomed in the entertainment, healthcare and manufacturing industries \cite{Kim2016IsVR}.
Impressive advances were achieved in visual and audio quality, with audio spatialization being now common place and human-eye resolution about the be reached \cite{Varjo}. However, although VR experiences are often praised as "fully immersive", bodily sensations are still largely missing, while they are crucial for several aspects of user experience \cite{reiner2004role}.

The perception of self-motion, in particular, can be induced visually but relies on multisensory cues, and notably on vestibular cues, i.e. acceleration and motion applied to the head \cite{britton2019vestibular}.
Although visual cues are usually more precise for perceiving displacement, in some cases they can be dominated by vestibular cues \cite{harris2000visual} \cite{fetsch2009dynamic}.

This perceptual discrepancy leads to cybersickness effects as well as user disinvestment (in particular for whole-body displacement such as vehicle driving, flying, falling, etc), both of which are major issues in many VR applications.
As a result, VR content producers tend to adapt to those technological limitations instead of fitting the best possible user experience. 
The most stunning example is probably the use of teleportation as a relatively standard locomotion system, despite huge design efforts for other metaphors with better realism and ecological agency \cite{boletsis2017new}.
In order to address this lack of physical sensations, over the last years both researchers and industrialists proposed a large number of haptic VR peripherals (see \cite{wang2019multimodal} for a review). 

Yet only a subset of them addressed motion sensations.
If the common approach used in amusement parks is to actuate the seats with cumbersome systems, multisensory integration can also be leveraged to get rich effects from lightweight solutions \cite{danieau2012hapseat}\cite{nilsson2012haptically}.

\begin{figure}[tb]
 \centering
 \includegraphics[width=\columnwidth]{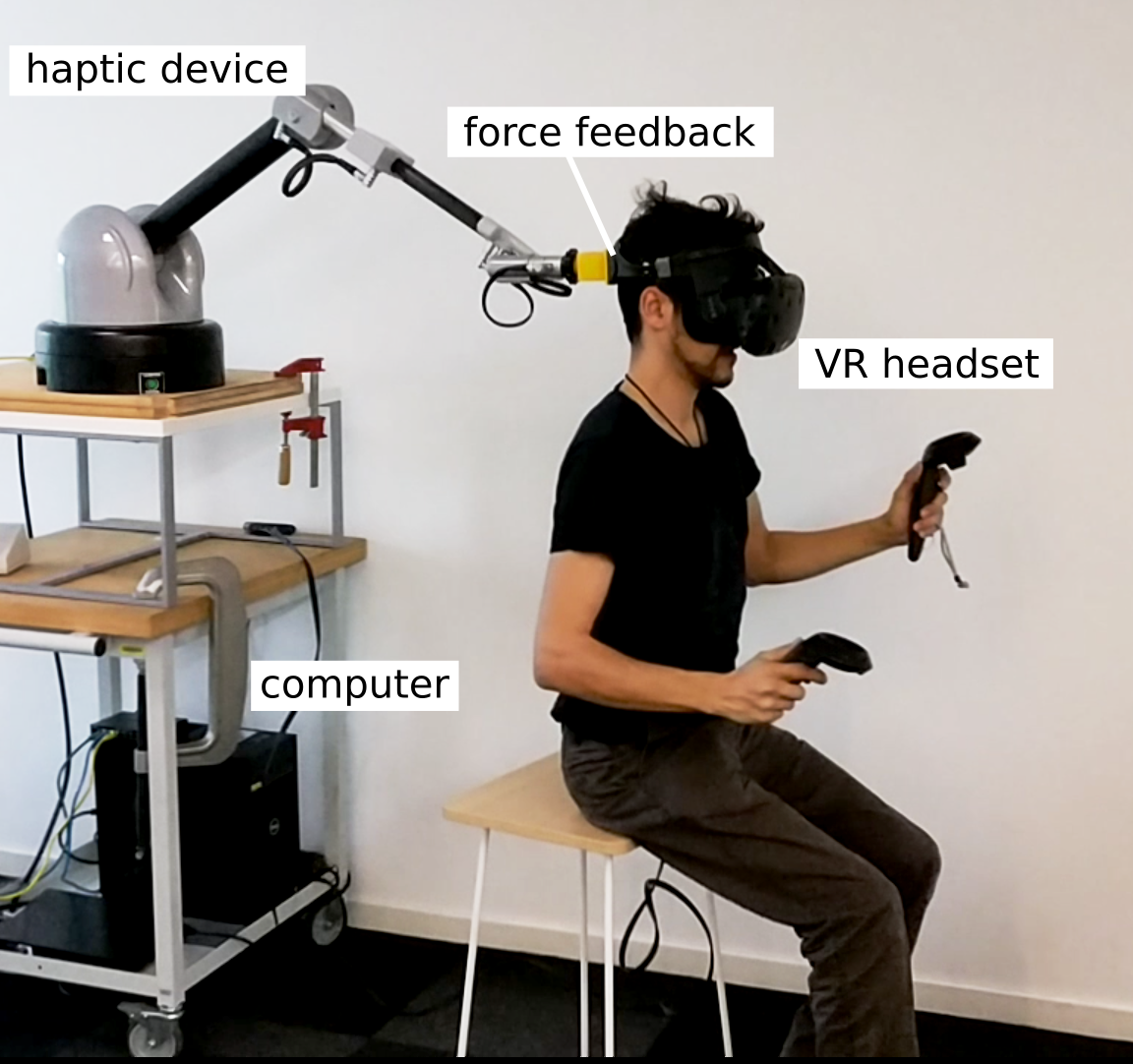}
 \caption{The Kinesthetic HMD: head-based force feedback enhancing self-motion sensations.}
 \label{fig:system}
\end{figure}

In this short paper, we propose a novel approach called the "Kinesthetic HMD" to enhance self-motion sensations in immersive VR applications.
It consists in actuating a head-mounted display (HMD) with force feedback in order to provide rich and versatile self-motion effects.
Similarly to a motion platform, our approach provides motion cues to the user, but at the level of his/her head, so to simulate whole-body accelerations and enhance illusory self-motion.
In contrast with actuated seats, the Kinesthetic HMD stimulates the vestibular system rather than the whole body.

In the remainder of this paper, we first present related work on VR peripherals providing motion sensations. 
Then, the Kinesthetic HMD approach is introduced and its hardware, software and safety design is discussed.
We demonstrate a flight simulator use case with a proof-of-concept prototype featuring a high-end grounded 6-DoF (degrees of freedom) haptic arm.
Then we present a user study focusing on the haptic enhancement of visual self-motion sensations, before finally discussing our experimental results and possible future works.

\section{Related work}

Although ``vection" (illusory self-motion) was historically considered as a visual illusion, it can be defined more broadly as a subjective conscious experience of self-motion" \cite{palmisano2015future}.
Its correlations with presence and immersion are studied, yet their relationship remain unclear \cite{prothero1995foreground} \cite{riecke2005scene}\cite{valjamae2006vibrotactile}.
It was found that exposure to vection in VR modulates vestibular processing \cite{gallagher2019vection}.

The vestibular system provides information about the angular rotation speed and linear acceleration of the head in space, which are crucial for self-motion estimation \cite{cullen2012vestibular}.
The relationship between visual and vestibular contributions to motion sensations is still not fully understood \cite{britton2019vestibular}, and they seem to be dynamically reweighted \cite{fetsch2009dynamic}.
Yet it can be said that the visual system is specialized for position and velocity estimation, while the vestibular system is optimized for acceleration processing.

This explains why the visual motion provided by a HMD, despite its ability to evoke illusory displacement, remains "incomplete" and not sufficient to provide compelling inertial sensations.
In order to get physical sensations, there is a need for physical stimulation.
Haptic devices provide kinesthetic and/or tactile stimulation. 
The vast majority of haptic solutions for VR are either wearable, holdable or grounded, and stimulate the finger or the hand in order to reproduce contact mechanics, material properties and physics of digital interaction (see \cite{wang2019multimodal} for a review).
Much less addressed self-motion sensations.

One approach to induce vection is to stimulate a large part of the skin, usually by integrating vibrators in a chair \cite{soave2020preliminary}.
Yet the most common technological solution is to provide actual motion to the seat: moving seats are usually found in amusement arcades for racing games and in theme parks for so-called 4D cinemas \cite{4dfilm}.
Those systems being usually cumbersome and expensive, and yet limited by a restricted amplitude, several researchers proposed simplified versions relying on illusions or sensory substitution \cite{rietzler2018vrspinning}\cite{teng2019aarnio}.
In particular, Danieau et al.\ proposed to affix multiple force-feedback devices to a chair, stimulating specific parts of the body (hands and head), in order to provide inexpensive 6-DoF motion effects \cite{danieau2012hapseat}.

Applying haptic stimulation on specific regions of the body (feet, hand or head) is indeed a promising approach to achieve rich effects with a minimalist setup. 
It was shown that a haptic stimulation of the feet could induce self-motion sensations while standing still in various virtual reality scenes \cite{nilsson2012haptically}, and modulates vection even in a seated position \cite{farkhatdinov2013vibrotactile}\cite{kruijff2016your}.
L\'ecuyer et al.\ investigated the perception improvement of visual turns by reproducing the turn angle (or its opposite) through a haptic handle \cite{lecuyer2004can}.
Ouarti et al.\ used a similar setup to show that a haptic feedback in the hands could improve duration and occurrence of visually-induced illusory self-motion for linear and curved trajectories \cite{ouarti2014haptic}.
Bouyer et al.\ extended this approach to an interactive video game context \cite{bouyer2017inducing}.


HMD-embedded haptics is a quite recent research topic.
Peng et al.\ showed that step-synchronized vibrotactile stimuli on the head could significantly reduce cybersickness \cite{peng2020walkinvibe}, while Wolf et al.\ proposed to combine vibrotactile and thermal feedback inside the HMD for an increased presence \cite{wolf2019faceon}.
Gugenheimer et al.\ attached flywheels to an Oculus Rift DK2 to generate torque feedback on the head \cite{gugenheimer2016gyrovr}.
A major drawback of this solution is the lack of transparency, as when the flywheel turns faster it builds up inertia against user's movements.
Kon et al.\ suggested to leverage the intriguing so-called "hanger reflex" effect, which is an involuntary head rotation arising from a specific pressure distribution, in order to provide various illusory forces \cite{kon2017hangerover}.
However this technique does not allow precise force rendering as it relies on muscular reflex that is likely to vary a lot among individuals.
Chang et al.\ proposed a pulley-based mechanism to produce normal force on the HMD \cite{chang2018facepush}.
This system could improve immersion in boxing or swimming simulations, but did not aim at generating self-motion sensations.
Finally, Wang et al.\ integrated skin-stretch modules inside a HTC Vive Pro to provide haptic feedback on the user's face \cite{wang2019masque}.
In a motorcycle racing simulation context, their system would simulate weight, bumping transients, inertial turns, and wind pressure.
However, to our knowledge, grounded force feedback was never applied to a VR headset.

\section{The Kinesthetic HMD approach}\label{"hhmd"}

\subsection{Concept}
The Kinesthetic HMD is a novel approach to enhance self-motion sensations in virtual reality with head-based force feedback.
It can be viewed as an augmentation of the HMD, adding precise forces and displacement to images and sound provided to the head.
The generated vestibular and proprioceptive cues emphase the visual motion to produce stronger and more compelling sensations of begin accelerated, although the user, standing or seated, remains in the same place.

The force feedback should be congruent with visual motion, for instance being proportional to ego-acceleration.
Yet the rendering algorithm can be adapted to the locomotion context.
When the user walks virtually, the haptic device can simulate head movements in compliance with the walking pace.
When the user gets in virtual car and drives, the force feedback can push or pull to simulate acceleration and braking.

The Kinesthetic HMD approach requires at least four components: a force-feedback device, a VR headset, a head clamping part transmitting the forces, and a software platform for VR content and haptic rendering.
Such a system implies hardware, software as well as safety considerations, detailed in the reminder of this section.

\subsection{Hardware considerations}
Applying force feedback to the head raises a number of technical challenges, especially as producing vestibular cues requires steady directional forces, able to move the head relatively to the thorax in one to three directions.
First, the actuator should benefit of adequate backing.
Second, the system should be comfortable enough not to break immersion.
Lastly, the forces should be correctly transmitted to the head.

\textbf{Backing}
Most force feedback devices requires backing, either on the body (wearable) or on a fixed frame (grounded).
Backing on the shoulders (or any body part other than the head) is complicated because of their mobility relatively to the head.
Holdable devices, which could be totally embedded on a helmet, come with significant limitations like motion impediment and saturation (reaction wheels and weight shifting devices), or buzzing (asymmetric vibrations).
Moreover their force capability is directly limited by their moving mass, and therefore their carrying weight.

\textbf{Comfort}
The head is more sensitive than other body parts to noise, motion restriction or extra weight.
Therefore mechanical actuators should ideally be properly isolated, or departed from the point of application.
Also, the mechanical transmission should not present any risk of wrong movements for the neck.

\textbf{Point of application}
The force feedback should not only be produced, but also properly transmitted, and choosing the adequate clamping system can be an issue..
The clamping part should not slip on the head, yet it should remain comfortable.
Pressure should be limited and therefore area of contact maximized, but avoiding ears or any sensitive part.
Even with a capable device, producing a constant directional force on the head is challenging because when the head rotates the alignment with its mass center is easily lost, resulting in unwanted counter torques.


\subsection{Software considerations}
Designing the proper self-motion rendering algorithms for head-based force feedbacl is not straight-forward for several reasons, detailed in this subsection.
First, the haptic stimulus should act as a metaphor, as we want the user to feel an illusory whole-body motion.
Second, the provided forces and displacement should ensure safety and all values are not acceptable.
The consequent filtering can impede realism, especially for real-time interactive content that is not known in advance.
Third, the effect of the force feedback might depend on the user's position, which should be taken in account in the rendering loop.

\textbf{Haptic rendering}
Several authors used velocity-based vestibular feedback to study vection \cite{riecke2006head} \cite{vailland2020vestibular}, but a comparative study suggested that acceleration is a better choice, producing stronger and more consistent illusory self-motion sensations \cite{ouarti2014haptic}.
Another advantage of acceleration over velocity is that the summed amplitude over time is lower, and thus is less prone to workspace limitation issues.
Among possible haptic metaphors, two were pointed out by previous works: the direct mode and the indirect mode \cite{lecuyer2004can} \cite{bouyer2017inducing}.
In the direct mode, the haptic feedback is proportional to visual acceleration: this means if the virtual vehicle speeds up, the haptic device will push forwards, simulating physical acceleration. In the indirect mode, the haptic feedback is inverted: when the virtual vehicle speeds up the haptic device will pull backwards, simulating the physical body displacement.

\textbf{Interactivity}
Just like the visual content, the haptic effects can be fully interactive (i.e. generated from user input), pre-recorded (i.e. experienced passively), or a mix of both (i.e. actively triggered or modulated).
Pre-recorded content can be analyzed and scaled to mitigate extreme values while preserving realism.
Fully interactive experiences might require thorough real-time filtering, depending on the content.
Trade-offs can be found with hybrid strategies, for instance triggering a pre-recorded sequence only when the user reaches the adequate part of the workspace.

\textbf{Motion filtering}
As our force feedback is expected to move the head, the user will lean (assuming they stay in the same position).
The leaning amplitude will be limited either by the user's anatomy or the haptic device's capabilities.
Just like with actuated seats, the rendering algorithms should avoid reaching the edges of the workspace, and eventually come back towards the center below perceptual thresholds, which can be achieved with so-called "washout filters" \cite{danieau2014kinesthetic}.
One potential issue in the context of head-based haptic rendering is to determine which perceptual thresholds (vestibular or proprioceptive) to consider.

\begin{figure}[tb]
 \centering
 \includegraphics[width=\columnwidth]{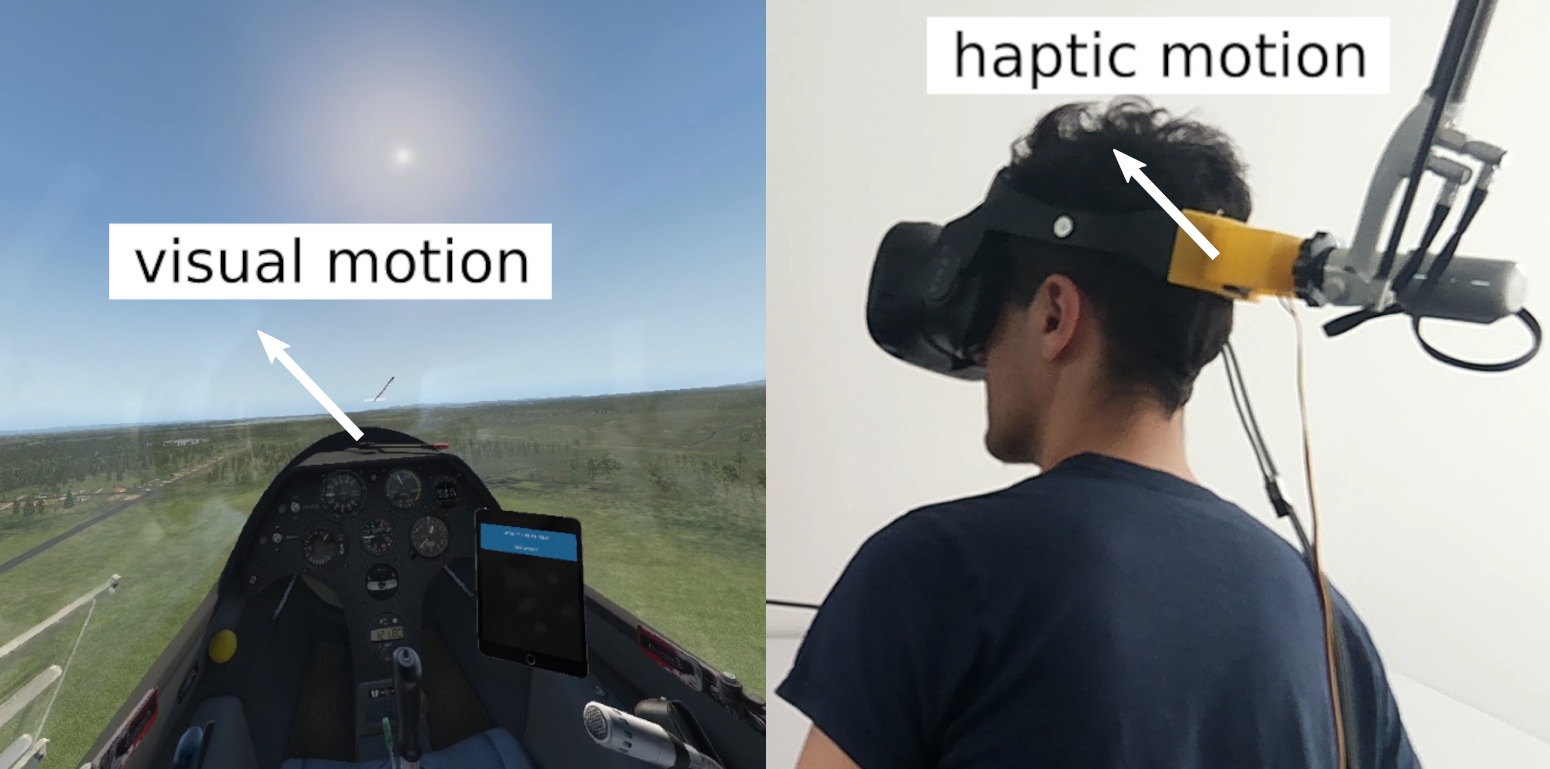}
 \caption{Flight simulation use case: the force feedback is inversely proportional to visual acceleration.}
 \label{fig:back}
\end{figure}


\subsection{Safety considerations}
Respecting anatomical limits and avoiding any risk of injury is of course a major requirement for any haptic setup, and even more when applied on the head.
Although defining general-case safe limits is tricky, an order of magnitude of $100N$ seem to be a low value both for neck muscular force in any direction for healthy adults \cite{lecompte2007biomecanique} and for neck loadings involved by common non-injurious physical activities \cite{funk2011head}.
Thus, values of one order of magnitude lower (dozens of $N$) might be considered for the maximal applied forces in healthy and warmed-up adults.
In addition to limiting the total applied force, we suggest to limit the applied jerk (to avoid false moves), to have a hardware kill switch, and to design pre-session tests to adapt the gain to each individual.
If the user is standing, the risk of unbalance and fall has also to be considered.

\section{Proof of concept}
In order to showcase our concept, we developed a prototype (see Figure~\ref{fig:system} for the detailed components) based on a high-end grounded haptic arm and tuned it for a flight simulation scenario.
That is, we wanted to provide the user strong and compelling sensations of acceleration, deceleration, gravitational roll and turbulence.
Also, we wanted our demo to be accessible without prior skills, so we chose a glider flight scenario, which requires minimal pilot skills.
The full scenario is detailed in the accompanying video.


\begin{figure}[!tb]
 \centering
 \includegraphics[width=\columnwidth]{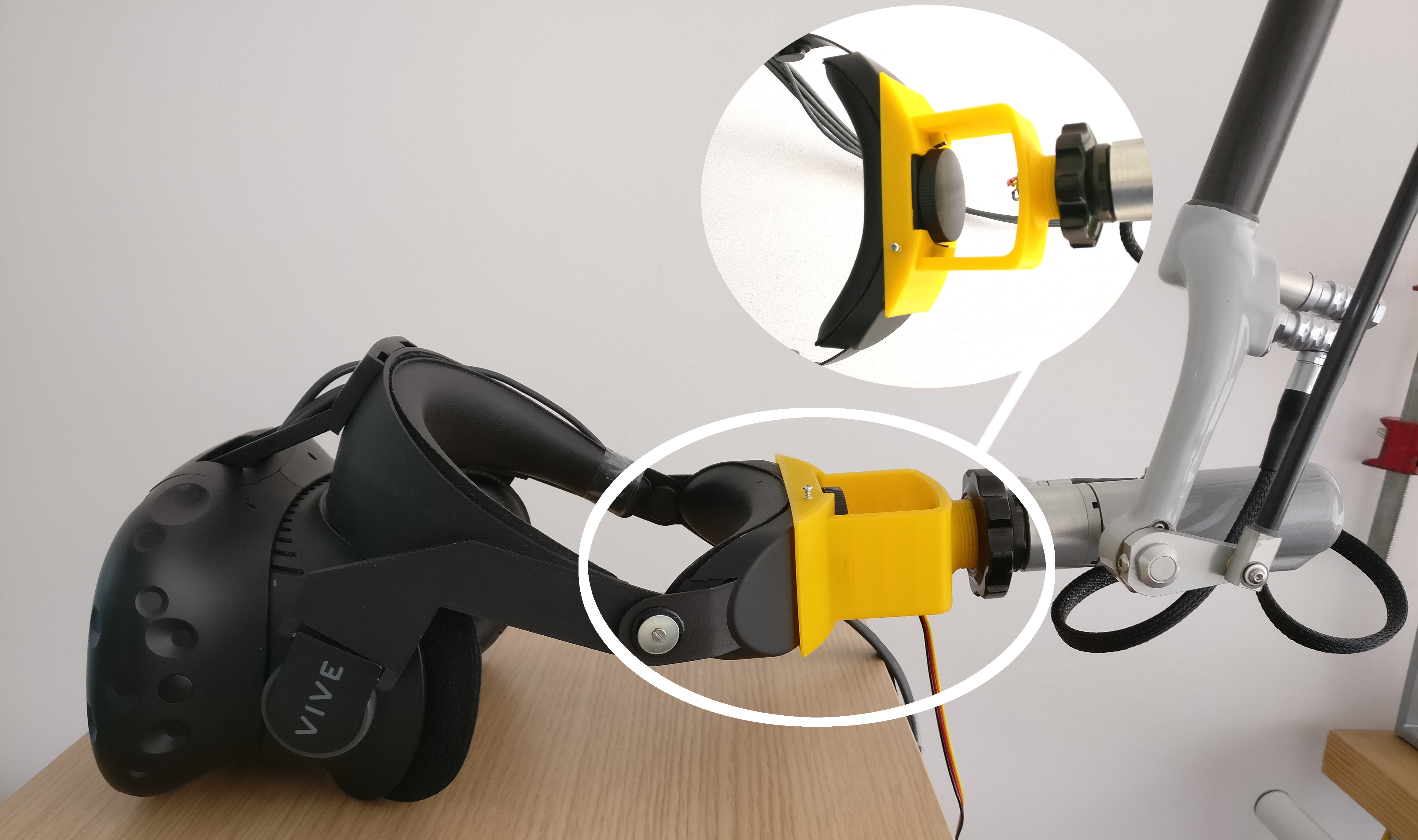}
 \caption{The rigid headband and the custom 3D-printed connector.}
 \label{fig:connector}
\end{figure}

\subsection{Hardware components}
We used a Virtuose 6D 35-45 haptic arm, which provides strong forces and torques in a workspace of about 1.3mx5mx1m.
We placed its base about 1m above the floor in order to have the user's head in the middle of the workspace.

By using a grounded device, we don't load the head with extra weight, and don't expose it to motor sound and vibrations.

For the HMD we used a HTC Vive headset with an unofficial "Rift S style" rigid headband found online.
We designed and 3D-printed a mechanical connector replacing the Virtuose handle, plugged at the back of the headband and secured with a screw (see Figure~\ref{fig:connector}).
The use of a 6-DoF haptic arm allows for cancelling counter torques when applying steady 3-DoF forces.


\subsection{Software components}
We built upon a glider flight scenario of a standard software flight simulator (X-Plane 11).
Flight simulation softwares output a series a real-time motion data, and notably acceleration values to be used for motion platforms.
We used a python script to receive the flight simulation acceleration data through UDP and to compute the haptic rendering.
The applied forces were negatively proportional to visual acceleration.
In order to cancel counter torques, when force modulus exceeded $1N$ the applied torques were simulating a virtual cylinder joint along the direction of the applied force.
When the forces were lower than $1N$ there were no applied torques in order to let the head free to move.

Another advantage of flight simulation softwares is that they allow for recording and replaying scenarios.
Therefore the user can choose between the intense experience of a fully interactive flight, or the comfort of a pre-recorded flight without the stress of manoeuvring the aircraft.
In both cases, the force feedback is rendered in real-time from the acceleration values output.

No "washout filter" was implemented as applied forces usually lasted only a few seconds and were limited to $10N$, so the user would not reach the edges of the workspace during our scenario.

\subsection{Safety considerations}
The Virtuose arm has a hardware security switch usually controlled by a proximity sensor integrated to the handle, so that force feedback is disabled whenever the handle is not held.
We replaced it with a mechanical button acting as a kill switch that could be held either by the user or the experimenter.
We limited the total applied force to $10N$.
The gain factor between visual acceleration and applied forces could also be refined upon user request to adjust experience intensity.

\section{User study}
In order to evaluate our system's ability to enhance visually-induced self-motion sensations, we conducted a user study comparing the motion sensations evoked by visual and visuo-haptic displacement stimuli.
On the basis of Ouarti et al.'s results, we chose to focus on acceleration-based haptic rendering, and also to evaluate both direct mode and indirect mode, as Ouarti et al.\ suggested that half the people would prefer one and half the people the other \cite{ouarti2014haptic}.

\subsection{Hypotheses and objectives}
In the context of virtual reality applications, we are not only interested in strengthening motion sensations, but also making them more compelling. 
That is, providing an illusory self-motion experience rather than a "scrolling landscape" feeling. 
In other words, we want the motion to be felt as more egocentric (self-motion rather than landscape motion) as well as more intense (with vivid bodily sensations).
Comfort should also be considered, as visuo-haptic discrepancy can produce immersion-breaking awkwardness if they are not congruent.

Therefore we designed several displacement stimuli and evaluated the sensations they produced as follow:
\begin{itemize}
    \item \textit{Relative motion}: am I moving in a fixed an environment or am I at rest watching a moving environment?
    \item \textit{Acceleration}: do I have bodily sensations similar to being in a moving vehicle with eyes closed?
    \item \textit{Comfort}: how pleasant or unpleasant is the motion experience?
\end{itemize}

We designed our study to test the following hypotheses:
\begin{itemize}
    \item \textbf{H1}: visuo-haptic stimuli induce more egocentric motion sensations than a similar visual stimulus
    \item \textbf{H2}: visuo-haptic stimuli induce stronger bodily sensations than a similar visual stimulus 
    \item \textbf{H3}: some participants would have higher ratings (in relative motion, acceleration and/or comfort) for the direct mode, and some other for the indirect mode
\end{itemize}
 
\begin{figure}[tb]
 \centering
 \includegraphics[width=\columnwidth]{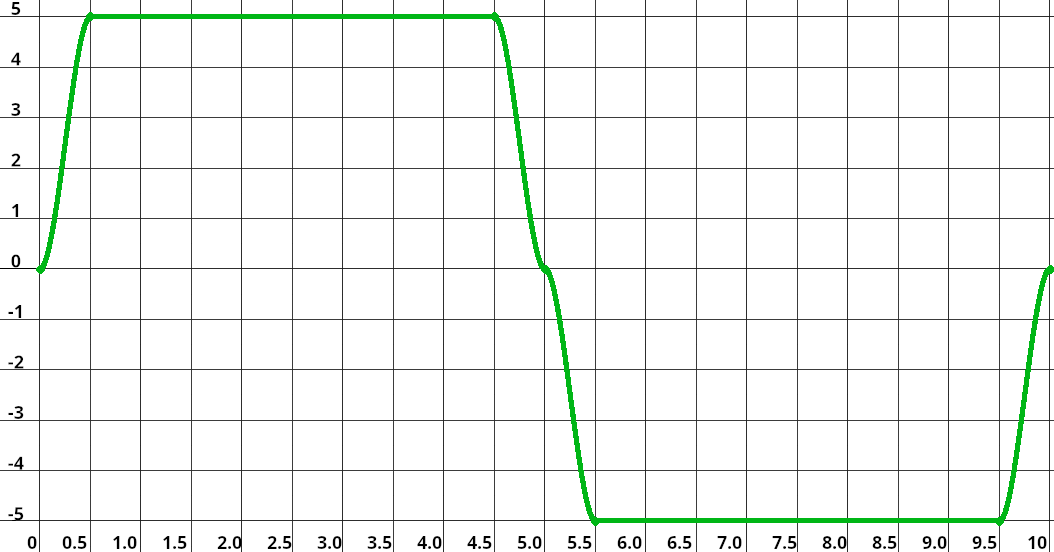}
 \caption{The temporal acceleration pattern used for the stimuli. The curve shows acceleration ($m.s^{-2}$) over time ($s$).}
 \label{fig:acc_pattern}
\end{figure}

\subsection{Stimuli}
The visual context was designed so as to be as neutral as possible.
The virtual environment was an empty space filled with a random spatial distribution of 20000 white cubes, vanishing in a black fog at about 70 meters distance. In order to mitigate the lack of embodiment and eventual vertigo symptoms, a white 3mx4m rectangle was used as a symbolic ground.

The displacement stimulus was based on a double step acceleration pattern (see Figure~\ref{fig:acc_pattern}): a forward acceleration for 5 seconds followed by a deceleration of 5 seconds. In order to avoid transient effects, those two steps were eased with 0.5sec long sinusoidal curves, so the acceleration magnitude was of a maximal of 5 $m.s^{-2}$ for 4 seconds.

The visual stimulus was identical for all conditions: the cubes were accelerated according to the previously depicted pattern, creating an ambiguous relative displacement between them and the participant and ground.
The three experimental conditions were as follow: 
%

\begin{itemize}
    \item \textbf{H{\_}NONE}: no force feedback
    \item \textbf{H{\_}DIRECT}: force feedback is proportional to the acceleration pattern: pushing during acceleration, pulling during deceleration
    \item \textbf{H{\_}INDIRECT}: the force feedback is proportional to the opposite of the acceleration pattern
\end{itemize} 

In both H{\_}DIRECT and H{\_}INDIRECT conditions, the haptic arm maintained head orientation constant during the stimulus to avoid any lever arm effect.

\subsection{Participants}
17 participants (2 females, age 23-54, mean=35.3, SD=10.7) volunteered for the experiment. 
They were all recruited in the research center, had corrected-to-normal vision, and no balance disorder history.
They all signed an informed consent form prior to participating to the experiment.
In order to prevent cybersickness symptoms, participants were asked to hydrate well before participating, and were free to have a break anytime during the experiment. 
One extra participant could not finish the experiment because of cybersickness symptoms, thus their incomplete data was removed from the results.

\subsection{Experimental design}
Each participant would attend a single session of about 50mn consisting in three phases: an introduction phase, an exploration phase, and an experimental phase.

In the introduction phase, the participant read and signed an informed consent form conforming to the declaration of Helsinki. 
In order to clarify the vocabulary used during the experiment, the participant was orally given examples of different motion sensations related from a displacement stimulus: estimated traveled distance, relative motion (which of me or the landscape is actually moving ?), and "acceleration", that is the non-visual sensation of being moved 
that can be felt when closing your eyes in a moving vehicle. 

\begin{figure}[b]
 \centering
 \includegraphics[width=\columnwidth]{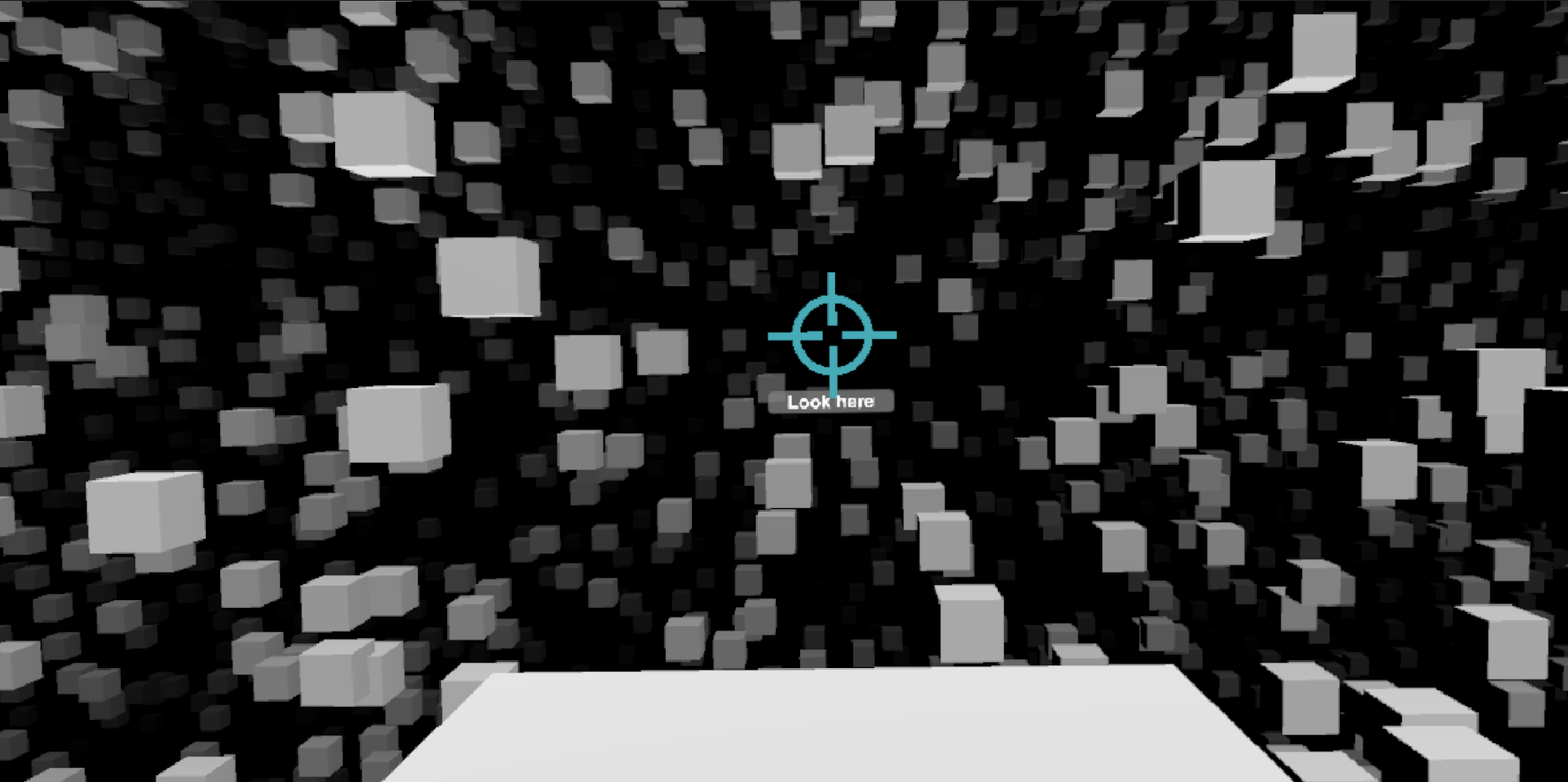}
 \caption{The visual stimulus used in the experiment.}
 \label{fig:wait}
\end{figure}

Then, the participant was introduced to the protocol through the exploration phase.
The participant sat on a stool and adjusted the HMD for a correct vision. 
The experimenter progressively explained the experimental conduct by running 8 trials (4 H{\_}NONE, 2 H{\_}DIRECT, 2 H{\_}INDIRECT) and answering any question (except about the nature of the stimuli). 
In order to avoid surprise effects, the haptic arm was not connected to the HMD before the third trial (the first visuo-haptic stimulus).
On one of the visuo-haptic trial, the participant was asked to perform a "security break exercise": they would say a safeword any time during the stimulus, after which the experimenter immediately released the kill switch to disable haptic feedback and cancel the trial.
The participant's answers were not recorded during the introduction phase.
After the eighth trial, the participant was offered to have a break, and the experimenter made sure they had no remaining question about the protocol or the task, before moving to the experimental phase.

\begin{figure}[t]
 \centering
 \includegraphics[width=\columnwidth]{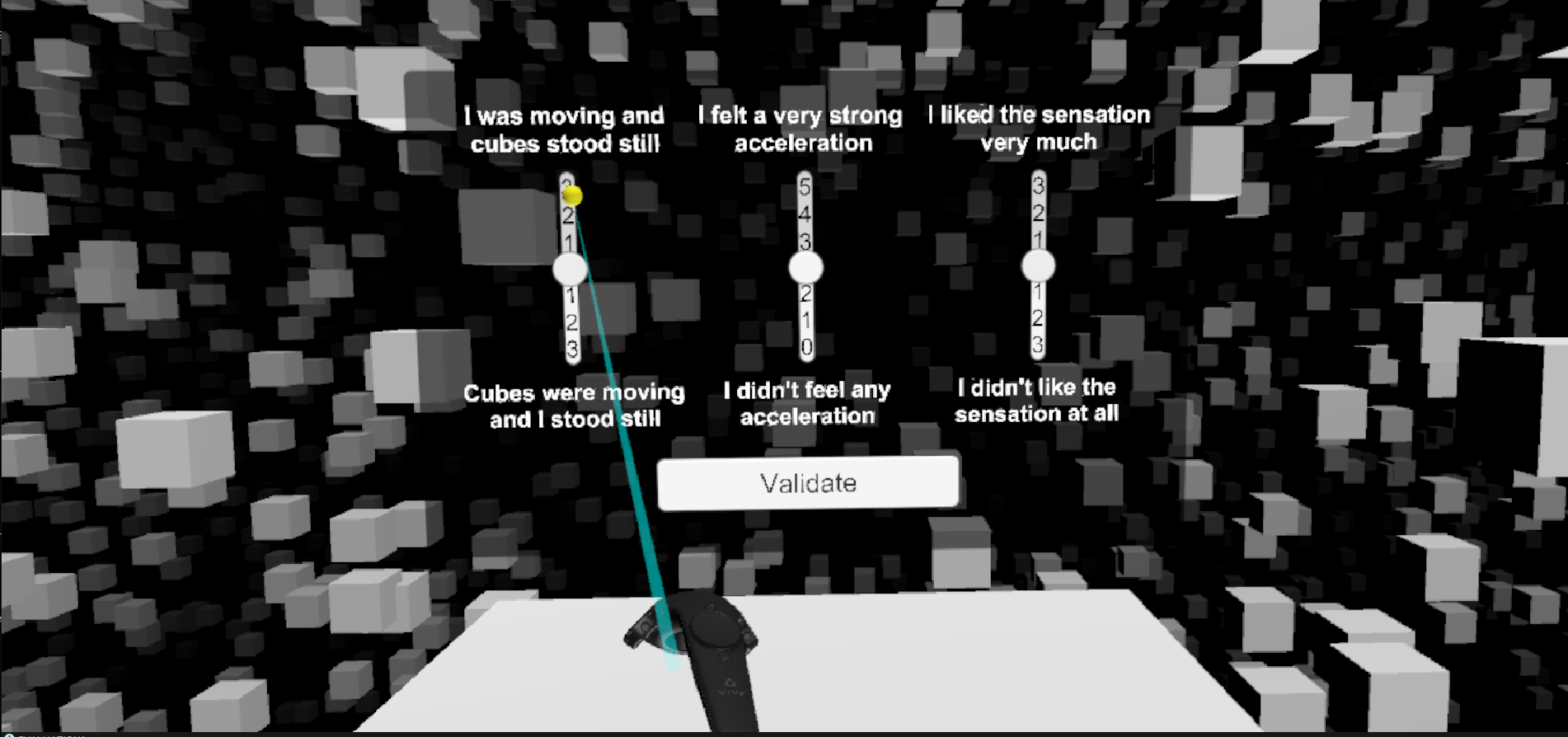}
 \caption{User rating during the experiment.}
 \label{fig:questions}
\end{figure}

The experimental phase consisted in a randomized block of 30 trials (10 repetitions of each three conditions). 
All the trials followed the same structure.
First, the participant validated the launch of the trial. Then, a visual target was displayed forward for 1.5 seconds (see Figure~\ref{fig:wait}). 
Then the stimulus was played for 10 seconds. 
Finally, the participant rated the stimulus on \textit{Relative Motion}, \textit{Acceleration} and \textit{Comfort} scales (see Figure~\ref{fig:questions}).

Participants were asked to keep their gaze towards the
direction indicated by the target at the beginning of the trial. 
They were asked to maintain their gaze direction and simply remain comfortable if the haptic stimulus would modify their posture.

\subsection{Results}
\begin{figure}[tb]
 \centering
 \includegraphics[width=\columnwidth]{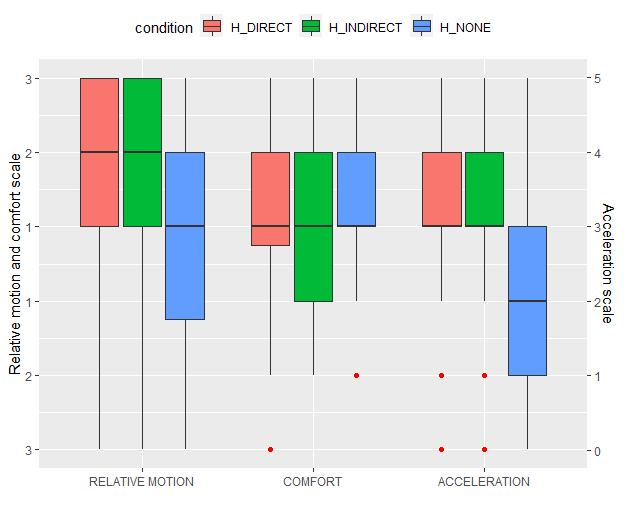}
 \caption{Ratings distributions across all participants. Red dots are outliers.}
 \label{fig:all_grouped}
\end{figure}

Participants' ratings for are shown on Figure~\ref{fig:all_grouped}.
\textit{Relative Motion} was mostly rated positive (egocentric motion) in the H{\_}DIRECT and H{\_}INDIRECT conditions, while answers were mixed for the H{\_}NONE condition.
\textit{Comfort} ratings were mostly weakly positive for the H{\_}DIRECT and H{\_}NONE condition, and mixed for the H{\_}INDIRECT condition.
\textit{Acceleration} was mostly rated above 3/5 in the H{\_}DIRECT and H{\_}INDIRECT conditions, while mostly under 3/5 in the H{\_}NONE condition.

For each of the three variables, as the normality assumption was violated (Shapiro–Wilk’s normality test, \(p < 0.05\)), a Friedman test was conducted independently followed by a post-hoc Wilcoxon-signed ranks test with a Bonferroni correction.
For \textit{Relative Motion} and \textit{Acceleration}, significant differences (\(p < 0.001\)) were found between H{\_}NONE and H{\_}DIRECT and between H{\_}NONE and H{\_}INDIRECT.
For \textit{Comfort}, a significant difference (\(p < 0.001\)) was found between H{\_}NONE and H{\_}INDIRECT only.
We didn't find any effect of age, gender or previous experience of virtual reality on any of the three variables.

Taken together our result validate both \textbf{H1} and \textbf{H2}, with significantly higher ratings of \textit{Relative Motion} and \textit{Acceleration} for visuo-haptic stimuli compared the to the visual-only stimulus.
Surprisingly, the rating distributions for both H{\_}DIRECT and H{\_}INDIRECT were very similar, and contrary to \textbf{H3} we could not split the results in two populations preferring one or the other type of haptic feedback, as suggested by previous work \cite{ouarti2014haptic}.
The only noticeable difference was the \textit{Comfort} ratings, slightly more negative for the H{\_}INDIRECT condition, leading to significant differences with the H{\_}NONE condition, which ratings were mostly positive (almost never judged as uncomfortable).

\section{General Discussion}
\subsection{Approach}
The informal reactions to our flight simulator prototype confirmed that the visuo-haptic experience was much more compelling than the visual-only one.
After having tried the haptic-enhanced scenario, the visual-only scenario felt "empty" or "flat", and also more prone to dizziness.
It seemed that both low-frequency and high frequency haptic feedback contributed strongly to the quality of experience: the roaring take-off would lack of power without the slow and strong pull, but would also lack of realism without the swift turbulences.

The rigid headband can be used separately from the haptic arm, but also from the HMD, which means we could use it with other visual conditions, from a simple screen to video-projected space or augmented reality glasses.
A quick fastener on various haptic devices could allow for a versatile usage.

Depending on the application the displacement might be active or passive, and the user can be seated or standing. 
The design of head-based haptic feedback for active displacement might lead to novel locomotion techniques, improving user performance and/or quality of experience.
When standing, the workspace can be limited depending on the chosen haptic device, and hardware should also be adapted to imbalance and falling issues.

The lack of bodily sensations is a major issue for virtual reality, limiting user quality of experience and agency, and increasing cybersickness occurrence.
Our approach opens thrilling possibilities to apply haptic cinematography principles, as proposed by Danieau et al.\ \cite{danieau2016introducing}.
Beyond the realism of physical simulations, creators can make use of non-diegetetic effects to increase dramatic intensity, and direct attention with staging or anticipation techniques.
More generally, haptics could be seen as a media by itself, able to improve storytelling and make better narratives, just like cinema sound design.

\subsection{User Study}
Taken together the results of our study show that, as expected, the visuo-haptic stimuli induced more vivid and more compelling self-motion sensations than a similar visual-only stimulus.
Those enhanced motion sensations might arise from vestibular and/or proprioceptive cues.
As the participants could resist more or less to the haptic feedback to maintain their posture, their leaning amplitude could vary a lot between the stimuli.
However we could not find, formally or informally, a clear difference in the accompanying sensations.
On top of that there was a significant difference in amplitude between individuals depending on their weight (which varied with a factor of two between participants).
Yet even weighted individuals with no visible leaning might answer with high self-motion ratings.
Further research should clarify the respective contributions of vestibular and proprioceptive cues.

The vestibular haptic cues might also alter the estimations travel distance, as suggested by previous work \cite{harris2000visual}.
We initially planned to include this task in our study, but the pilot tests revealed an surprisingly high cognitive load for making both distance and velocity/acceleration estimations.
After providing a traveled distance estimate, the participants could hardly make assessments about relative motion or non-visual sensations and vice-versa, as if those two kinds of task had conflicting short memory processes.
The perception of displacement given by visual and visuo-haptic stimuli remains to be studied.

\subsection{Future Work}
Head-based force feedback opens exciting possibilities in terms of haptic rendering, training simulation and virtual reality quality of experience.
It also comes with various application-related challenges to address in future work.

Our prototype based on a 6-DoF haptic arm was able to provide stable forces in any of the three directions. 
Yet both visual and vestibular perceptual thresholds are anisotropic \cite{crane2014human}, thus directionality could be taken in account when elaborating head-based visuo-haptic stimuli, and the psychophysical thresholds could be investigated.

High individual variance, habituation and attentional phenomenons as well as influence from various top-down factors \cite{riecke2005topdown} need to be taken in account to achieve the rendering of bodily sensations.
Control laws could be adjusted depending on the user as well as the virtual environment context, like getting in a virtual vehicle or flying.
Those adjustment might be achieved with offline tests to estimate user sensibility, as well as tuned in real-time depending on user's state, similarly to motion platform washout algorithms.
The role of habituation and attention could also be further studied.

For specific scenarios, smaller haptic devices could be used.
For instance, for a flight simulation altitude control exercise, the force feedback might only simulate gravitational tilt: when the plane pitches up, the user is pulled backwards, and when the plane is pointing down the user is pushed forwards.

\section{Conclusion}
In this short paper, we proposed a novel approach to make self-motion sensations stronger and more compelling in virtual reality applications, by means of head-based force feedback.
The provided vestibular and proprioceptive cues enhance the perception of self-motion, making it more egocentric and more bodily, while preserving user's comfort.

We discussed the technical and scientific challenges raised by head-based force feedback and demonstrated a flight simulator use case with a proof-of-concept prototype based on a high-end grounded haptic arm.
We evaluated our system with a user study focused on the motion sensations provided by visuo-haptic displacement stimuli.
Our results showed that the visuo-haptic rendering induced more vivid and more egocentric sensations of self-motion than a similar visual-only rendering.

\ifCLASSOPTIONcaptionsoff
  \newpage
\fi

\begin{IEEEbiography}{Antoine Costes}
is a research engineer at Inria Rennes, France. He obtained a PhD in Computer Science from INSA Rennes in 2018, within a collaboration between the Immersive Lab of InterDigital company (formerly Technicolor R\&I) and the Hybrid research team at Inria Rennes. His research topics include pseudo-haptic rendering and human perception in immersive environments. He prototypes connected objects, interactive systems and haptic actuators for both scientific and artistic purposes.
\end{IEEEbiography}

\begin{IEEEbiography}{Anatole L\'ecuyer}
 is a senior researcher and head of the Hybrid team, Inria Rennes, France. He is currently an associate editor of the IEEE Transactions on Visualization and Computer Graphics, Frontiers in Virtual Environments, and Presence journals, as well as formerly of the ACM Transactions on Applied Perception and the International Journal of Humanities and Cultural Studies. He was a program chair of the IEEE Virtual Reality Conference(2015-2016) and the IEEE Symposium on 3D User Interfaces (2012-2013).
\end{IEEEbiography}





\end{document}